# Intermolecular CT excitons enable nanosecond excited-state lifetimes in NIR-absorbing non-fullerene acceptors for efficient organic solar cells

Xian-Kai Chen[1,2,3,10,*], Christopher C.S. Chan[4,5,10], Sudhi Mahadevan[2,10], Yu Guo[4], Guichuan Zhang[6], He Yan[5], Kam Sing Wong[5,*], Hin-Lap Yip[2,7,8], Jean-Luc Bredas[9], Sai Wing Tsang[2,*], Philip C.Y. Chow[4,*]

1. Department of Chemistry, City University of Hong Kong, Hong Kong, P. R. China
2. Department of Material Science and Engineering, City University of Hong Kong, Hong Kong, P. R. China
3. Hong Kong Institute for Advanced Study (HKIAS), City University of Hong Kong, Hong Kong, P. R. China
4. Department of Mechanical Engineering, The University of Hong Kong, Pokfulam, Hong Kong, P. R. China
5. Department of Physics, Department of Chemistry and William Mong Institute of Nano Science and Technology, Hong Kong University of Science and Technology, Clear Water Bay, Hong Kong, P. R. China
6. State Key Laboratory of Luminescent Materials and Devices, Institute of Polymer Optoelectronic Materials and Devices, School of Materials Science and Engineering, South China University of Technology, 381 Wushan Road, 510640 Guangzhou, P. R. China.
7. School of Energy and Environment, City University of Hong Kong, Hong Kong, P. R. China
8. Hong Kong Institute for Clean Energy, City University of Hong Kong, Hong Kong, P. R. China
9. Department of Chemistry and Biochemistry, The University of Arizona, Tucson, AZ 85721-0088, USA
10. These authors contributed equally

**Abstract**

State-of-the-art Y6-type molecular acceptors exhibit nanosecond excited-state lifetimes despite their low optical gaps (~1.4 eV), thus allowing organic solar cells (OSCs) to achieve highly efficient charge generation with extended near-infrared (NIR) absorption range (up to ~1000 nm). However, the precise molecular-level mechanism that enables low-energy excited states in Y6-type acceptors to achieve nanosecond lifetimes has remained elusive. Here, we demonstrate that the distinct packing of Y6 molecules in film leads to a strong intermolecular charge-transfer (iCT) character of the lowest excited state in Y6 aggregates, which is absent in other low-gap acceptors such as ITIC. Due to strong electronic couplings between the adjacent Y6 molecules, the iCT-exciton energies are greatly reduced by up to ~0.25 eV with respect to excitons formed in separated molecules. Importantly, despite their low energies, the iCT excitons have reduced non-adiabatic electron-vibration couplings with the electronic ground state, thus suppressing non-radiative recombination and allowing Y6 to overcome the well-known energy gap law. Our results reveal the fundamental relationship between molecular packing and nanosecond excited-state lifetimes in NIR-absorbing Y6-type acceptors underlying the outstanding performance of Y6-based OSCs.

**Introduction**

The recent development of Y6-type non-fullerene acceptors (Y6-NFAs) has led to significant improvement of the power conversion efficiencies of bulk-heterojunction organic solar cells (OSCs), which now approach 19% in single-junction devices.[1] These Y6-NFAs have an acceptor-donor-acceptor-donor-acceptor (A-DA'D-A) molecular structure comprising a fused thienothienopyrrolo-thienothienoindole (TTP-TTI) core and two electron-withdrawing end units.[2] This molecular structure distinguishes Y6-NFA from the earlier generation of NFA molecules, which adopt an A-D-A structure with a fused-ring indaceno[1,2-b:5,6-b']dithiophene (IDT) core and two electron-withdrawing end units.[3] Notable examples include ITIC and IT-4F, and we refer to these IDT-based acceptors as I-NFAs.[4] Y6-NFAs have now replaced both I-NFAs and fullerenes as the ubiquitous acceptor materials in efficient OSC devices. Thus, gaining a fundamental understanding of the electronic properties of Y6-NFAs behind their remarkable device performances is crucial for further development of OSC research.

Previous reports have shed light on the relationships between excited-state properties of Y6-NFAs and OSC performance.[5,6,7,8,9,10,11,12,13,14,15,16,17,18] Brabec and co-workers used a Boltzmann stationary-state equilibrium model to demonstrate a direct correlation between lifetime of the lowest singlet excited state ($S_1$) and free charge generation efficiency in NFA-based OSC devices.[9] For OSC blends with negligible driving force ($\Delta E_{S1-CT_{D/A}}$) for exciton dissociation at the donor-acceptor (D/A) interface (which is defined as the energy offset between the $S_1$ state localized on D or A and the interfacial charge-transfer ($CT_{D/A}$) state), free-charge generation at the D/A interface takes up to tens to hundreds of picoseconds to complete at room temperature.[5,13] The $S_1$-state lifetime in the neat Y6 film is over 1 ns, which ensures that, at thermal equilibrium, localized electron-hole pairs can effectively separate into free charge carriers with near-unity quantum efficiency.[19,20] In contrast, model I-NFA systems such as ITIC and IT-4F show much reduced $S_1$-state lifetime, which is consistent with the inferior device performance.[9,10] Also, the $S_1$-state energy of Y6 film ($E_{S1} \sim 1.40$ eV) is much lower compared to the model I-NFAs (typically ~1.65 eV), thus allowing enhanced near-infrared absorption (up to ~1000 nm) that in turn leads to higher device photocurrents.[2,10] Therefore, both the nanosecond $S_1$-state lifetime and the low $E_{S1}$ are crucial excited-state properties of Y6-NFAs underpinning their high performances in OSC devices.

However, the precise molecular-level mechanism that enables Y6-NFAs to achieve longer $S_1$-state lifetimes than those in I-NFAs has remained elusive. Since $E_{S1}$ is much lower in the Y6-NFAs, faster non-radiative recombination rates ($k_{nr}$) and therefore shorter $S_1$-state lifetime are expected according to the energy gap law for exciton-vibration relaxation (*i.e.*, $k_{nr} \propto \exp(-E_{S1})$; see also **Equation 1** in **Methods** for details).[21,22,23] This suggests that Y6-NFAs exhibit very different excited-state properties compared to I-NFAs. Earlier reports by some of us and others have shown that the Y6-NFA aggregates exhibit a distinct molecular packing in both the neat and blend films that are not found in the I-NFA aggregates,[10,24] but the role of such molecular packing in the $S_1$-state energy and nanosecond lifetime is not understood to date.

In this report, we use a combination of optical spectroscopy measurements and quantum-chemical calculations to study the fundamental excited-state properties of Y6-NFA monomers and aggregates. We demonstrate that strong electronic couplings between adjacent molecules in Y6-NFA aggregates lead to the formation of delocalized $S_1$ states with greatly reduced energy (up to ~ 0.25 eV) with respect to the localized $S_1$ states formed in monomers, thus enabling strong NIR absorption. Interestingly, despite the much reduced $S_1$-state energy, the $S_1$-state lifetimes and $k_{nr}$ in Y6-NFA aggregates are nearly unchanged with respect to those in the monomers. Our results show that the distinct molecular packing of Y6-NFA aggregates gives rise to a strong intermolecular charge-transfer (iCT) character of the $S_1$ states. The iCT-excitons have reduced non-adiabatic electron-vibration couplings with the electronic ground state ($S_0$), which compensates the effect of the reduced energy on the non-radiative recombination rate. We propose that this iCT-exciton-driven mechanism enables low-energy, NIR-absorbing $S_1$ states in Y6-NFA aggregates to overcome the well-known energy gap law and achieve nanosecond lifetimes, as required for efficient charge generation in OSCs.

**Materials**

Our study involves four model Y6-NFA molecules (Y6, Y7, BTP-eC9 and Y6-BO), and four well-established I-NFA molecules are also studied for the sake of comparison (ITIC, IT-4F, IT-M and IEICO-4F)[2,3,25]. **Figure 1** shows their chemical structures and their $E_{S1}$ in neat spin-cast films. For consistency, we determine $E_{S1}$ by considering the intersection of the steady-state absorption and PL spectra (taken using calibrated detectors; see **Methods** and **SI** for details).[26] Time-resolved PL measurements show a mono-exponential decay for all systems (**Figure 1b**). The fitted lifetimes, which correspond to the $S_1$ lifetime ($\tau$) of the system, are plotted against $E_{S1}$ in **Figure 1c**. We find that all of the studied Y6-NFAs exhibit a significantly extended $S_1$-state lifetime (~1.5-1.8 ns) compared to the I-NFAs (~0.3-0.7 ns). This indicates that, despite their lower $E_{S1}$, the formation of the nanosecond-lived and emissive $S_1$ states is indeed a general feature in Y6-NFAs.

**Absorption and emission properties**

We have studied absorption and PL properties of the NFA molecules in dispersed states (found in dilute solutions) and aggregated states (found in neat spin-cast films). **Figures 2a,b** show the steady-state absorption/PL spectra of Y6 and ITIC, respectively (see **SI** for results on other

NFAs). A spectral red-shift is observed for both systems in films with respect to solutions, reflecting a lowering of $E_{S1}$ (dashed lines). For ITIC, the $E_{S1}$ is reduced by about 0.12 eV (from ~1.79 to 1.67 eV), which is likely due to increased electronic polarization caused by weak molecular aggregations. For Y6, we find a greater $E_{S1}$ drop as the molecules form aggregates in films (by ~0.25 eV, from ~1.63 to 1.38 eV). Similar observations are found for the other NFAs, thus indicating stronger intermolecular interactions in the $S_1$ state properties of Y6-NFAs than in I-NFAs. By comparing absorption and PL spectra, we find that both Y6-NFA and I-NFA neat films show similar Stokes shift of about 0.2 eV, indicating that they have similar structural relaxation ($\lambda$ in **Equation 1** in **Methods**).

**Figure 2c** shows the PL decay kinetics of Y6 and ITIC in dilute solutions and neat films, showing similar $S_1$ lifetimes in dispersed and aggregated states for both Y6 (~1.5 ns) and ITIC (~0.35 ns). Using a calibrated integrating sphere (see **Methods**), we measured the PL quantum yields (PLQY) of various NFAs in both dilute solutions and in spin-cast films (**Figure 2d**). For I-NFAs (ITIC and IT-4F), we find similar PLQY values in dilute solutions and neat films. The similarity in $S_1$ lifetime and PLQY measured for dispersed and aggregated I-NFAs indicates that the Frenkel excitons localized on the isolated molecules dominate the $S_1$ state in both cases, thus reflecting a weak intermolecular interaction in their excited-state properties. This is consistent with the loose molecular packing between the end groups found in I-NFA films.[27] In contrast, for all four Y6-NFAs, we observe significantly different PLQY values in dilute solution (~15-17%) and neat films (~5-7%). This indicates that the $S_1$ states formed in dispersed and aggregated Y6-NFA molecules are significantly different. Since dispersed Y6-NFA molecules show significantly higher PLQY than aggregates (about 3 times higher), the similar $S_1$ lifetime indicates a large difference in the radiative recombination rate ($k_r$) between the $S_1$ states in dispersed/aggregated molecules. The reason is that the total $S_1$ state decay rate $(1/\tau)$ is equal to the sum of $k_r$ and $k_{nr}$ (non-radiative recombination rate) and PLQY is given by $k_r/(k_r + k_{nr})$.[28] The calculated values for $k_r$ and $k_{nr}$ are shown in **Figure 2e** and **Table S1** in **SI**. For I-NFAs, we observe similar $k_r$ and $k_{nr}$ in both solution and film ($k_{nr} \sim 2 \times 10^9$ s$^{-1}$ and $k_r \sim 5 \times 10^7$ s$^{-1}$). In comparison, dispersed Y6-NFAs have high $k_r$ ($\sim 10^8$ s$^{-1}$) and lower $k_{nr}$ ($\sim 6 \times 10^8$ s$^{-1}$), giving rise to the high PLQY. When Y6-NFA molecules start aggregating, we find that $k_r$ is reduced by a factor of about three while $k_{nr}$ remains largely unchanged. We will return to this point in the discussion below.

**Electroabsorption spectroscopy**

The absorption and PL results clearly suggest that, for Y6-NFAs, the nature of $S_1$ states are different in dispersed and aggregated molecules. Electroabsorption (EA) spectroscopy is a powerful technique where the spectral characteristics depends on the excitonic properties of materials. In EA as described by the Stark effect, the change in excitation energy ($\Delta E_{S1}$) under a perturbation of an electrical field ($F$) depends on the differences in molecular polarizability ($\Delta p$) and dipole moment ($\Delta\mu$) between the involved states of the optical transition., i.e. $\Delta E_{S1}(F) = -\Delta\mu F - \Delta p \frac{F^2}{2}$, where $\Delta E_{S1}$ is correlated with the change in optical transmission $\Delta T/T$ measured in device.[29] When Y6 molecules are dispersed in a matrix of an insulating polymer (polyvinylcarbazole or PVK) in thin films, the absorption spectrum is similar to that measured for dilute solutions at low concentrations, thus indicating that the molecules are sufficiently dispersed to mitigate the aggregation effects (see **Fig. S8** in **SI**). Upon sandwiching both PVK-dispersed and neat Y6 films between electrodes, we have measured the change in device optical transmission in response to an applied electric field due to the Stark effect (monitoring the second harmonic signal; see **Methods**). The resulting EA spectral responses of aggregated and dispersed Y6 molecules are shown in **Figures 3a,b**, while the corresponding results for ITIC are shown in **Figures 3c,d** for comparison. For ITIC, both dispersed and neat films show an EA response that largely resembles the first derivative of the absorption spectrum at the first optical transition (between the $S_1$ and $S_0$ states). In contrast, we find very different EA responses for dispersed and aggregated Y6 molecules. While the EA response of dispersed Y6 also resembles the first derivative of its absorption spectrum, in neat films the aggregated Y6 molecules exhibit an EA response that largely matches with the second derivative of its absorption near $E_{S1}$.

It is well-established that the second-harmonic EA response of organic conjugated semiconductors is given by the sum of the first and second derivatives of the absorption spectrum.[29,30,31] The first derivative term originates from the difference in polarizability ($\Delta p$) and the second derivative term scales with the change in dipole moment ($\Delta \mu$) between the states involved in the optical transition. Importantly, $\Delta \mu$ is the measure of the amount of iCT character associated with a transition. Therefore, for systems that form $S_1$ states with remarkable iCT character, the EA response near $E_{S1}$ is dominated by the second derivative component. This is indeed the case for the aggregated Y6 molecules in neat films. When the Y6 molecules are dispersed in insulating PVK (1:9 weight ratio), the EA response becomes closer to the first derivative component, thus reflecting an increasing population of $S_1$ states localized on single molecules with weak iCT character (*i.e.*, Frenkel excitons). For I-NFA molecules, the EA response is largely dominated by the first derivative component for both dispersed and neat films, which is consistent with weak intermolecular interactions and the localized nature of the $S_1$ states even in molecular-aggregates. The above EA results provide a valuable insight into the increase of iCT characteristics with the distinct molecular packing in Y6 aggregates which is absent in I-NFA aggregates.

## Quantum-chemical calculations

To further characterize the excited-state properties of the Y6-NFA molecular aggregates, we turned to quantum-chemical calculations on the Y6 monomer and dimers (see the **Methods** for computational details). From snap-shots of the molecular-dynamics (MD) simulations reported previously by some of us, we found that the three Y6 dimer configurations exist in the single crystal, in the neat Y6 film, and in the Y6:polymer blend film.[10,27] Here, we refer to these Y6 dimer configurations as Dimer 1, 2, and 3, respectively, and we examine their electronic properties in detail. The calculated Natural Transition Orbitals (NTO) describing the excitation character of the $S_1$ states in the Y6 monomer and the three Y6 dimers are shown in **Figure 4a**. For the Y6 monomer, the NTO hole and electron wavefunctions of the $S_1$ state have a large spatial overlap ($O_{h/e} \sim 0.65$), representative of a Frenkel exciton nature. This leads to a large fluorescence oscillator strength ($f_{flo} \sim 2.0$), which is consistent with the fast $k_r$ (since $k_r \propto f_{flo}$) observed in Y6 dilute solution (**Figure 2e**). On the other hand, for the Y6 dimers, the calculated NTO hole and electron wavefunctions show the formation of the $S_1$ state with iCT character, which leads to a lowering of both $O_{h/e}$ (0.65 down to 0.40) and $f_{flo}$ (2.0 down to 0.2) in the Y6 dimers compared to the monomer. Due to the strong intermolecular electronic couplings in the dimer configurations that we calculated previously,[10] the calculated $S_1$ energies of the Y6 dimers are lowered with respect to that in the Y6 monomer. In particular, in Dimer-2 configuration, the face-to-face packing between the electron-withdrawing end-group of one Y6 molecule and the electron-donating core moiety of the adjacent Y6 molecule gives rise to a remarkable iCT character, leading to a large energy downshift of ~0.16 eV. Note that the calculated $S_1$ energies in the Y6 dimers do not account for the change in electronic polarization between solution and film, which could lead to additional energy down-shifting (as observed for the I-NFAs). Therefore, the computational results are consistent with the above experimental observation showing a spectral red-shift of ~0.25 eV in $E_{S1}$ energy measured in Y6-NFA films relative to dilute solutions (**Figure 2d**).

The remarkable iCT excitation character of the $S_1$ state in Y6 molecular aggregates matches the experimental observations described above. As $k_r \propto f_{\text{flo}}$, this rationalizes the reduced $k_r$ value (about three-fold) measured in pristine films of Y6-NFAs compared to dilute solutions (**Figure 2e**). Interestingly, the measured $k_{nr}$ values for both films and solutions of Y6-NFAs are nearly equivalent, despite the significant lowering in $E_{S1}$ energies in films (~0.25 eV) which should lead to a faster $k_{nr}$ according to the energy gap law (see **Equation 1** in **Methods**). At this stage, it is useful to recall that $k_{nr} \propto NAC^2 \cdot \exp(-E_{S1})$, where NAC denotes the non-adiabatic electron-vibration couplings between the $S_1$ and $S_0$ states. **Figure 4b** shows the calculated NAC values as a function of the vibration frequency for the Y6 monomer and the three Y6 dimers. Interestingly, we find that the maximum NAC values (~ 990 cm$^{-1}$) in all of the Y6 dimers are reduced compared to the Y6 monomer (~1100 cm$^{-1}$). We also evaluated the frequency-averaged NAC for Dimer-1, 2, and 3 (90 cm$^{-1}$, 72 cm$^{-1}$, and 70 cm$^{-1}$, respectively); these values are smaller than that (110 cm$^{-1}$) for the monomer (see **Equations 2-5** in **Methods**). Since NAC scales positively with $O_{h/e}$ in the $S_1$ state, the reduced NAC in dimers can be attributed to the remarkable iCT character of the $S_1$ states.

While the reduced $E_{S1}$ energy (from 1.65 to 1.4 eV) should lead to an increase in $k_{nr}$, since $k_{nr}$ scales positively with NAC$^2$, the iCT-exciton formation leads to a drop in $k_{nr}$ by a factor of ~0.42 to ~0.66 due to the reduced frequency-averaged NAC value (~0.83 when comparing the maximum NAC values). Thus, the reduced NAC values of the iCT-excitons compensates the effect of the reduced $E_{S1}$ energy on the non-radiative $S_1$-to-$S_0$ transition rate. This result is fully consistent with the similar $k_{nr}$ values measured for both Y6-NFA films and dilute solutions (**Figure 2e**).

**Discussion**

The combination of our experimental and theoretical investigations has thus revealed the fundamental excited-state properties of the Y6-NFA aggregates that enable them to exhibit simultaneously strong NIR absorption (up to ~1000 nm), long $S_1$ lifetime (> 1 ns), and high emission efficiency (PLQY > 5%). As described in the **Introduction** section, all of these excited-state properties are needed to ensure high charge-generation efficiency in OSC blends with a negligible $\Delta E_{S1-CT_{D/A}}$. The key implications of our results are as follows:

(1) The strong electronic couplings between the adjacent Y6 molecules lead to a substantial lowering (~0.25 eV) in the $S_1$-state energy of the Y6 aggregate with respect to the monomer. This feature opens up optical absorption deep in the NIR region (*i.e.,* up to ~ 1000 nm), thus enabling high device short-circuit photocurrent.

(2) Importantly, the iCT character of the lowest excited state in the Y6 aggregates serves to lower their NAC with the $S_0$ states, due to the reduced spatial overlap between the electron and hole wavefunctions. The effects on $k_{nr}$ of both the reduced $E_{S1}$ and NAC arising from the iCT-exciton formation compensates each other. Therefore, the similar $k_{nr}$ values (~ 6 x $10^8$ s$^{-1}$) are observed in both the pristine film and dilute solution of Y6-NFAs despite the substantial lowering in $E_{S1}$. As a result the low-energy, near-infrared absorbing $S_1$ states in Y6-NFA aggregates can achieve a long lifetime (> 1 ns), as required to promote efficient charge separation at the D/A interface.

(3) The large fluorescence oscillator strength of the Y6-NFA monomer leads to fast $k_r$ and high PLQY (~17% at ~800 nm) in solution. However, in Y6-NFA aggregates, the formation of the iCT-excitons causes a $k_r$ lowering and a near three-fold drop in PLQY (~6% at ~950 nm) compared to the monomers. This PLQY lowering in aggregates (found in D/A blends) is detrimental to the device open-circuit voltage as it induces a greater non-radiative voltage loss.[15,32] In the case of the hybridization of the aggregate iCT-exciton state and the more emissive monomer Frenkel-exciton state, optimizing $k_r$ of the monomer excitons would enhance the fluorescence oscillator strengths of the iCT-excitons via an intensity borrowing mechanism.[33] We envision this can be achieved via engineering of the molecular transition dipole; however, this requires further investigations and is beyond the scope of the present study.

(4) Finally, it is interesting to consider that further improving monomer emission and exploiting the iCT-exciton character of the Y6-NFA aggregates could also facilitate the realization of high-performance NIR organic light-emitting diodes.[23]

**Conclusion**

In summary, engineering long-lived and emissive excited states in NIR-absorbing organic semiconductors is needed to optimize OSC efficiency. However, this is a challenging task since exciton-vibration relaxation rate scales exponentially with decreasing excited-state energy, leading to fast non-radiative recombination rates. In this work, we have showed that the distinct molecular packing of Y6-NFA aggregates in films gives rise to the strong intermolecular CT (iCT) excitation character of the lowest excited state. Due to the strong electronic couplings between adjacent molecules in Y6-NFA aggregates, the iCT-exciton energies are greatly reduced (by up to ~0.25 eV) with respect to the excited states formed in separated molecules. Importantly, the formation of the iCT excitons plays a critical role in enabling Y6-NFAs to exhibit excited states with nanosecond lifetime and high NIR emissivity in aggregates. Within the framework of the energy gap law, an increased non-radiative recombination rate ($k_{nr}$) would be expected due to the reduced $E_{S1}$ energy. However, we found that the low-energy iCT-excitons on aggregates maintain a similar $k_{nr}$ as the excited states localized on monomers. The reason is that the iCT excitons have reduced non-adiabatic electron-vibration couplings with the $S_0$ states, thus enabling nanosecond lifetimes and high NIR emissivity. Although our study focuses on the aggregate excited states formed in the neat Y6-NFA films, previous reports have demonstrated that similar excited states exist in blended films of Y6 and wide-gap donor polymers.[10,18,27] Therefore, the iCT-exciton properties that we have uncovered here are directly applicable to Y6-NFA-based OSC devices. Overall, these excited-state properties enable state-of-the-art Y6-NFA-based OSCs to overcome the energy gap law for non-radiative recombination that applies to other NIR-absorbing OSC materials developed to date. In addition, our results may also provide useful design guidelines for the development of future NIR light-emitting and biosensing organic materials.

**Methods**

<u>Sample preparation</u>

ITIC, IT-4F, IT-M and IEICO-4F were purchased from Solarmer Inc (China). Y6, Y7, BTP-eC9 and Y6-BO were purchased from Derthon Optoelectronic Materials Science Technology Co. Ltd. (China). The materials were weighed and dissolved inside a nitrogen glovebox. The solutions were diluted to the desired concentration and sealed inside a quartz cuvette for optical measurements. For thin-film measurements, the solutions were spin-coated inside the glovebox onto precleaned quartz substrates to form a ~100 nm-thick thin-film. No thermal annealing was performed, and all film samples were encapsulated inside the glovebox to avoid degradation during measurements in air. Semi-transparent devices were fabricated for electroabsorption measurement in transmission mode. The devices had a general structure of ITO (120nm)/organic layer (150-200 nm)/Aluminium (15nm). The organic materials were dissolved in chloroform and stirred overnight before the spin-coating process. The solution was then spin-coated inside a nitrogen filled glovebox onto precleaned and UV-treated ITO glass substrates without further annealing. For the dispersed Y6 and ITIC films, controlled amount of polymer (polyvinylcarbazole or PVK) was added to the NFA solution to achieve certain weight ratio between the two materials. The samples were then transferred to a vacuum chamber (below $10^{-6}$ Torr) and thermally deposited with a 15 nm of aluminium as the top electrode to form semi-transparent devices. The thickness of the films was measured by a surface profilometer.

Absorption and emission spectroscopy

Steady-state absorption spectra were obtained using a Perkin Elmer Lambda 20 UV/VIS Spectrophotometer. For emission spectral measurements, we photoexcited the solution or thin-film samples near their absorption peak (refer to Figure S1 and S2), and the photoluminescence (PL) was collected by achromatic lenses and directed to a spectrometer that has been calibrated using a tungsten lamp. As described in the text, we determine the $E_{S1}$ by taking the intersection of the steady-state absorption and PL spectra. For time-resolved PL measurements, pulsed laser output either from a picosecond diode laser emitting at 640 nm (Edinburgh instruments EPL640) or a tunable Ti:sapphire oscillator (~700-1050 nm, Coherent Mira900) was used to photoexcite the sample. Excitation power was controlled with a variable ND filter and a calibrated laser power meter (Newport). The PL directed to a spectrometer (Acton SpectraPro275) equipped with time-correlated single photon counter system (Becker and Hickl SPC150 and PMC-100-20). An InGaAs single photon counter (MPD PDM-IR) was also used to verify the accuracy of the data. We ensured that excited state annihilations were minimized by using low solution concentrations (~$10^{-3}$ mg/ml) and excitation fluences (<0.05 $\mu J$ $cm^{-2}$). PLQY measurements were performed by photoexciting the solution or thin film samples at ~633 to 660 nm inside an integrating sphere (Labsphere) which is fiber-coupled to a calibrated spectrometer (Ocean Optics). Data acquisition and analysis were carried out following a previously reported protocol.[28] Additional spectroscopic data are found in the **SI**.

Electroabsorption spectroscopy

The electroabsorption measurements were conducted in transmission mode $\Delta T/T$.[28] Monochromatic light was generated using a Xenon lamp coupled with a monochromator equipped with various long pass filters to eliminate the $2^{nd}$ harmonic diffraction signal. The perturbation of electrical field was applied on the device by connecting the ITO and Al electrodes to a function generator (SRS DS360). An AC signal with a frequency of 1kHz and a typical electrical field of around $5 \times 10^5$ V/cm was applied on the sample. The monochromatic light beam was then focused inside the active area (3mm x 3mm) of the semi-transparent device (see above for sample preparation), where a calibrated silicon detector was used to measure the transmitted(existed) light intensity after passing through the device. The modulated signal from the silicon detector was amplified using a current pre-amplifier (Stanford Research Systems, SR570), which was then output to the lock-in amplifier (Stanford Research Systems, SR830) where the $2^{nd}$ harmonic signal with respect to the applied AC frequency where obtained. During the measurement, the samples were housed inside a vacuum cryostat (Oxford Instruments) with a base pressure below $10^{-5}$ Torr.

Quantum chemical calculations

Here, all the quantum-chemical calculations were carried out via the range-separated functional $\omega$B97XD and the 6-31 G(d,p) basis set.[34] Following our earlier investigations,[35] an iteration procedure was employed to non-empirically tune the $\omega$ parameter with the implicit consideration of the dielectric environment via the polarizable continuum model (PCM); the dielectric constant $\varepsilon$ was chosen to be 4.0, a representative value of organic semiconductor materials. All the excited-state properties of the Y6 monomer and dimers were examined at the time-dependent (TD) tuned-$\omega$B97XD/6-31 G(d,p) level of theory. NTO analyses were performed to characterize the characters of the $S_1$ states; the spatial overlaps between the density distributions of the hole and electron wavefunctions in the $S_1$ states were estimated with the Multiwfn code.[36] All the above TD-DFT calculations were performed using Gaussian 16 package.[37]

The non-radiative recombination rate of excited states in organic materials is expected to increase exponentially with reducing $E_{S1}$, as described quantitatively by:[23]

$$k_{nr} = \frac{(NAC)^2 \sqrt{2\pi}}{\hbar \sqrt{\omega E_{S1}}} exp \left\{ \frac{-E_{S1}}{\omega} \left[ ln \left( \frac{E_{S1}}{l\lambda} \right) - 1 \right] \right\} \tag{1}$$

where NAC is the non-adiabatic electronic couplings between $S_1$ and $S_0$ states, $\omega$ is the vibrational energy, $l$ is the number of vibrational modes that induce the nonradiative transition, and $\lambda$ is the reorganization energy of the promoting vibrational mode.

The non-adiabatic coupling ($\left\langle \Phi_{S_0} \left| \frac{\partial}{\partial Q_k} \right| \Phi_{S_1} \right\rangle$) between the $S_1$ and $S_0$ states for the $k$th vibrational normal mode can be expressed as[38,39]:

$$\left\langle \Phi_{S_0} \left| \frac{\partial}{\partial Q_k} \right| \Phi_{S_1} \right\rangle = \frac{\left\langle \Phi_{S_0}^0 \left| \frac{\partial V}{\partial Q_k} \right| \Phi_{S_1}^0 \right\rangle}{E\left(\Phi_{S_1}^0\right) - E\left(\Phi_{S_0}^0\right)} \tag{2}$$

where $V$ stands for the Coulomb interaction potential between electrons and nuclei; $\Phi_{S_1[S_0]}^0$ denotes the electronic eigenstate at the equilibrium position (marked by superscript "0") of the molecular structure in the ground state; $E\left(\Phi_{S_1[S_0]}^0\right)$ is the corresponding eigenenergy and $Q_k$, the coordinate of the $k$th normal mode. The numeration on the right hand side of Equation 2 writes:

$$\left\langle \Phi_{S_0}^0 \left| \frac{\partial V}{\partial Q_k} \right| \Phi_{S_1}^0 \right\rangle = -e \sum_{\sigma}^{atoms} Z_{\sigma} \left( \sum_j L_{\sigma,k}^j F_{S_1 \to S_0, \sigma}^j \right) \tag{3}$$

Here, $j = x, y$ or $z$; $Z_\sigma$ represents the nuclear charge; $r_\alpha^j$ and $R_\sigma^j$ are the Cartesian coordinates of electron $\alpha$ and nucleus $\sigma$, respectively; $L_{\sigma,k}^j = \frac{\partial R_\sigma^j}{\partial Q_k}$ denotes the atomic displacement of normal mode $k$; $F_{S_1 \to S_0,\sigma}^j$ represents the transition matrix element over the one-electron electric field operator at atomic center $\sigma$:

$$F_{S_1 \to S_0,\sigma}^j = \left\langle \Phi_{S_0}^0 \left| \sum_\alpha^{electrons} \frac{e(r_\alpha^j - R_\sigma^j)}{|\vec{r_\alpha} - \vec{R_\sigma}|^3} \right| \Phi_{S_1}^0 \right\rangle = \int d\vec{r} \, \rho_{S_1 S_0}^0(\vec{r}) \sum_\alpha^{electrons} \frac{e(r_\alpha^j - R_\sigma^j)}{|\vec{r_\alpha} - \vec{R_\sigma}|^3} \qquad (4)$$

where $\rho_{S_1 S_0}^0(\vec{r})$ is the electronic transition density at the equilibrium position. Equations (2) - (4) have been successfully applied to evaluate the non-adiabatic couplings between the excited state and ground state in organic emitters.[39] Equation (4) underlines that the matrix element is directly proportional to $\rho_{S_1 S_0}^0(\vec{r})$, *i.e.,* the spatial overlap between the hole and electron wavefunctions in the excited state.

The frequency-averaged NAC values (NAC$_{av}$) for the Y6 monomer and dimers are evaluated via:

$$NAC_{av} = \frac{\sum_i^n \hbar\omega_i NAC_i}{\sum_i^n \hbar\omega_i} \qquad (5)$$

where $NAC_i$ denotes the NAC value related to the $i$th vibration with energy $\hbar\omega_i$.

**Data Availability**

The authors declare that all relevant data are included in the paper and its Supplementary Information.

**Supplementary Information**

Supplementary Figs. 1 to 10 and Table 1.

**Acknowledgements**

X.-K.C. acknowledges the New Faculty Start-up Grant of the City University of Hong Kong (7200709 and 9610547). P.C.Y.C. acknowledges support from the Hong Kong Research Grant Council (16302520 and C6023-19G) and seed funding from the University of Hong Kong. G.Z. thanks the National Natural Science Foundation of China (No. 51903095), the Natural Science Foundation of Guangdong Province (No. 2021A1515010959) for financial support. K.S.W acknowledges support from the Hong Kong Research Grant Council (C6023-19G) and William Mong Institute of Nano Science and Technology (WMINST19SC04). S.-W. T. acknowledges the financial support from Hong Kong Research Grant Council (110303618) and CityU Strategy fund (7005643). The work at Arizona was funded by the Office of Navy Research, Award No. N00014-20-1-2110.

**Author contributions**

X.-K.C. and P.C.Y.C. conceived the project. X.-K.C. carried out all of the theoretical simulations. C.C.S.C., G.Y., K.S.W. and P.C.Y.C. carried out the absorption and emission experiments and analysis. S.M. and S.W.T. performed the electroabsorption measurements. G.Z. and H-L.Y. prepared samples. X.-K.C. and P.C.Y.C. wrote the manuscript. H.Y. and J-L.B. helped revise the manuscript. All authors discussed the results and commented on the final manuscript.

**Conflict of Interests**

The authors declare no conflict of interest

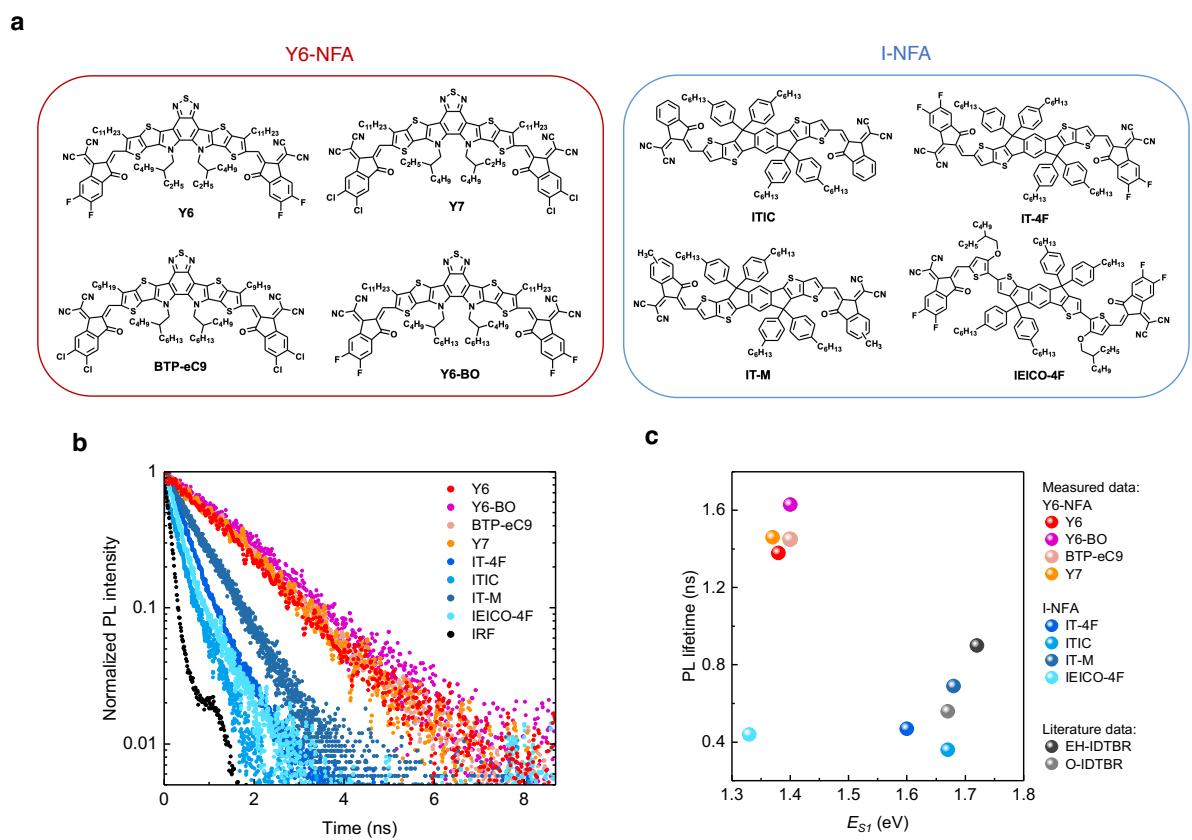

**Figure 1. Material characterizations.** (a) Chemical structures of the non-fullerene acceptors (NFA) considered in this study. (b) Semi-logarithmic plots showing time-resolved photoluminescence (PL) decay kinetics in a pristine spin-cast film of each NFA material, probed at the emission peak of the samples (see **Figures S1,2** for full spectra). Mono-exponential fit is used to extract the PL decay lifetime. IRF plots the response of the instrument. (c) Relationship between PL decay lifetime and the lowest singlet excited-state energy ($E_{S1}$) for film samples of these NFA materials. For consistency, we determine $E_{S1}$ by considering the intersection of the steady-state absorption and PL spectra (taken using calibrated detectors; see **Figure S1,2** and **Methods** for details).[26]

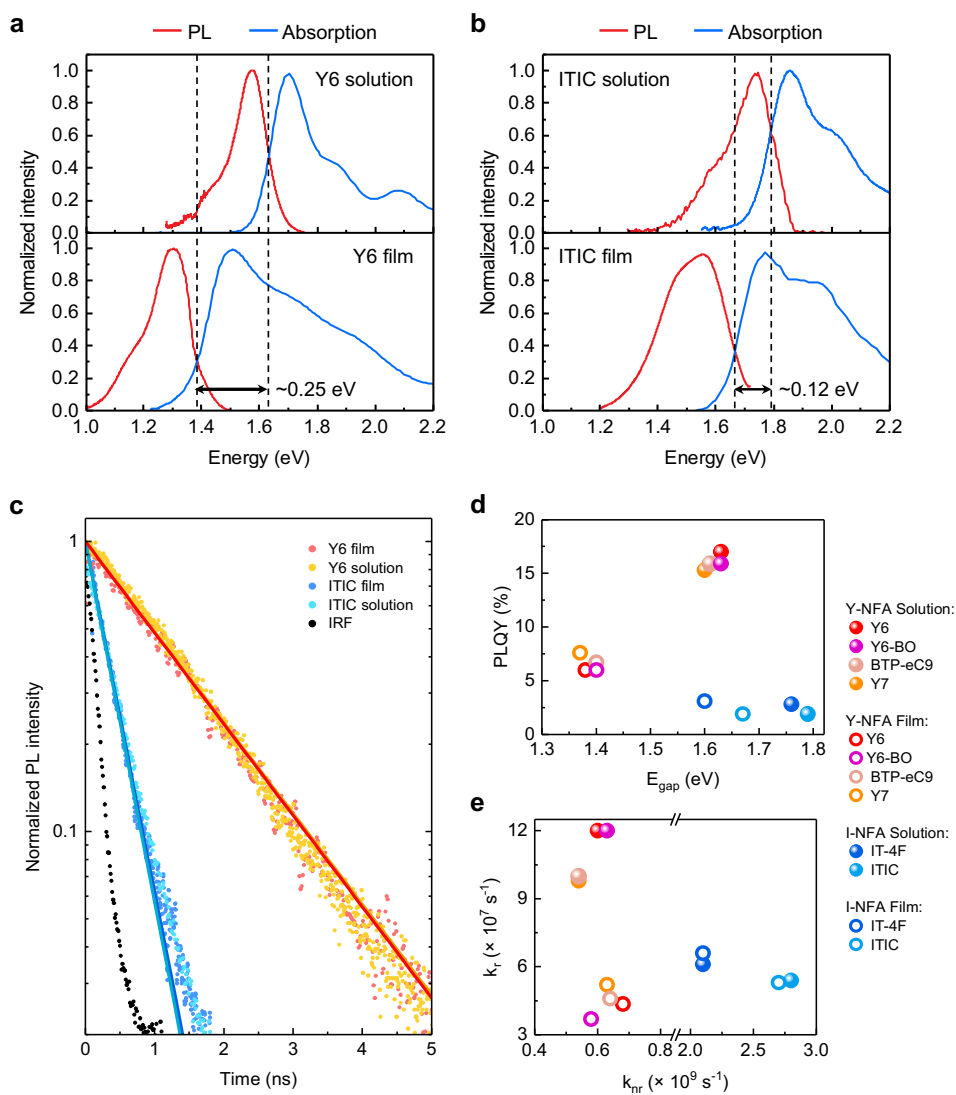

**Figure 2. Absorption and emission properties.** (a,b) Steady-state absorbance and photoluminescence (PL) spectra of Y6 and ITIC in dilute chlorobenzene solution ($10^{-3}$ mg/ml) and in neat spin-casted film on quartz substrates (see **Figure S1,2** in **SI** for other NFAs). (c) Semi-logarithmic plot showing time-resolved PL decay kinetics probed at the emission peak of the samples. The solid lines represent mono-exponential fits, and the decay time-constant corresponds to the lifetime of the first singlet excited state ($S_1$). IRF plots the response of the instrument. Additional PL decay results are shown in **Figure S3-S7** in **SI**. (d) PL quantum yield (PLQY) of various NFA molecules in dilute chlorobenzene solutions and in neat spin-casted films measured using a calibrated integrating sphere. (e) Radiative ($k_r$) and non-radiative recombination ($k_{nr}$) rates of each system determined from PLQY and PL decay measurements.

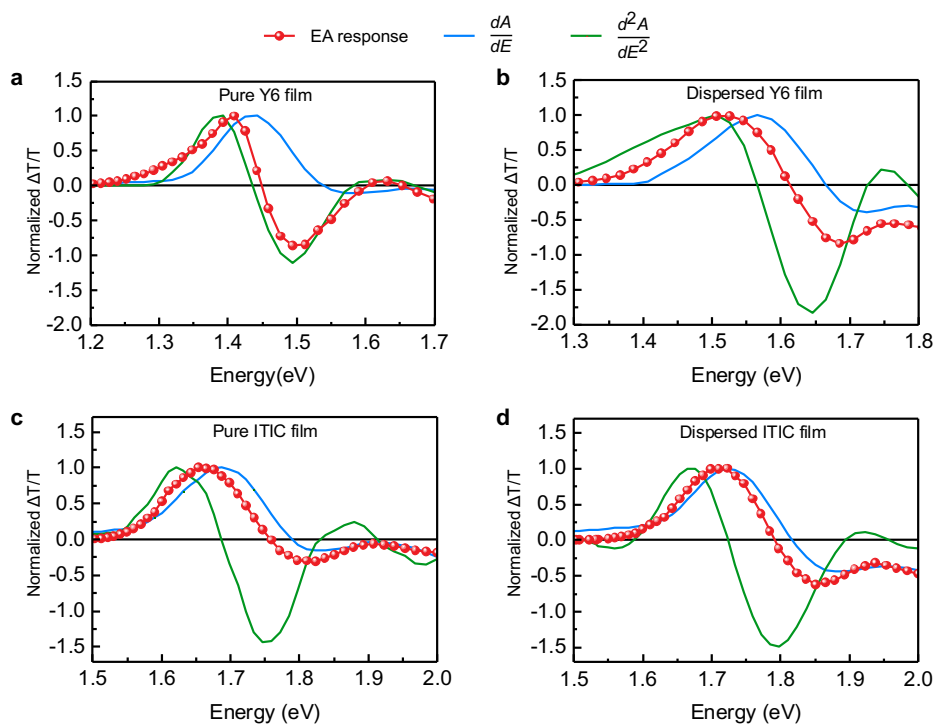

**Figure 3. Electroabsorption spectroscopy.** Dispersed and neat Y6 and ITIC film samples were sandwiched between semitransparent electrodes, and the normalized change in transmission ($\Delta T/T$) due to an applied electric field is monitored using a lock-in amplifier (see **Methods**). For dispersed Y6 and ITIC samples, the molecules were dispersed in an insulting polymer matrix (PVK) at 10% weight ratio (see concentration-dependent absorption in **Figure S8** in **SI**). The second harmonic electroabsorption response (red dots) are compared with the first (blue lines) and second (green lines) derivatives of the device absorption spectrum (unnormalized data are shown in **Figure S9** in SI).

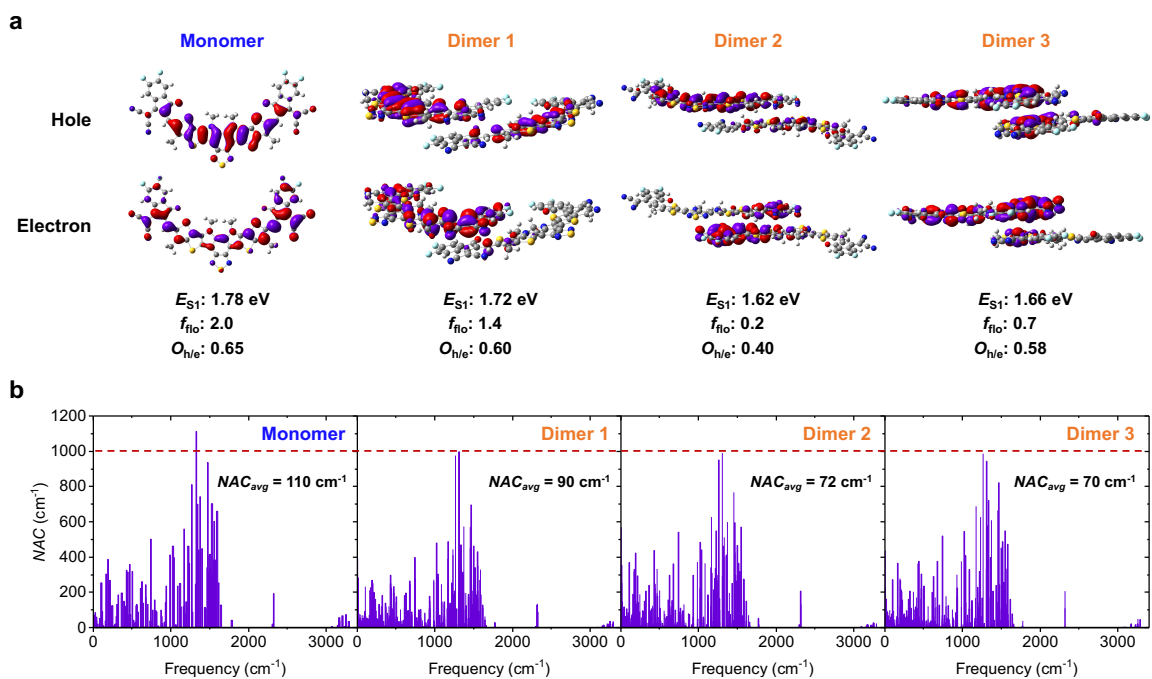

**Figure 4. Quantum-chemical calculation results.** (a) Natural transition orbitals (hole and electron) of the first singlet excited state (S₁) in the Y6 monomer and the three Y6 dimers (see **Figure S10** in **SI** for top view). S₁-state energies ($E_{S1}$), fluorescent oscillator strength ($f_{flu}$) and electron/hole spatial overlap ($O_{h/e}$) are summarized. (b) Non-adiabatic couplings (NAC) between the S₁ and ground states as a function of molecular vibrational frequencies for the Y6 monomer and dimers. The frequency-averaged NAC values for Y6 monomer, dimer-1, dimer-2 and dimer-3 are 110 cm⁻¹, 90 cm⁻¹, 72 cm⁻¹ and 70 cm⁻¹, respectively.

**Supplementary Information**

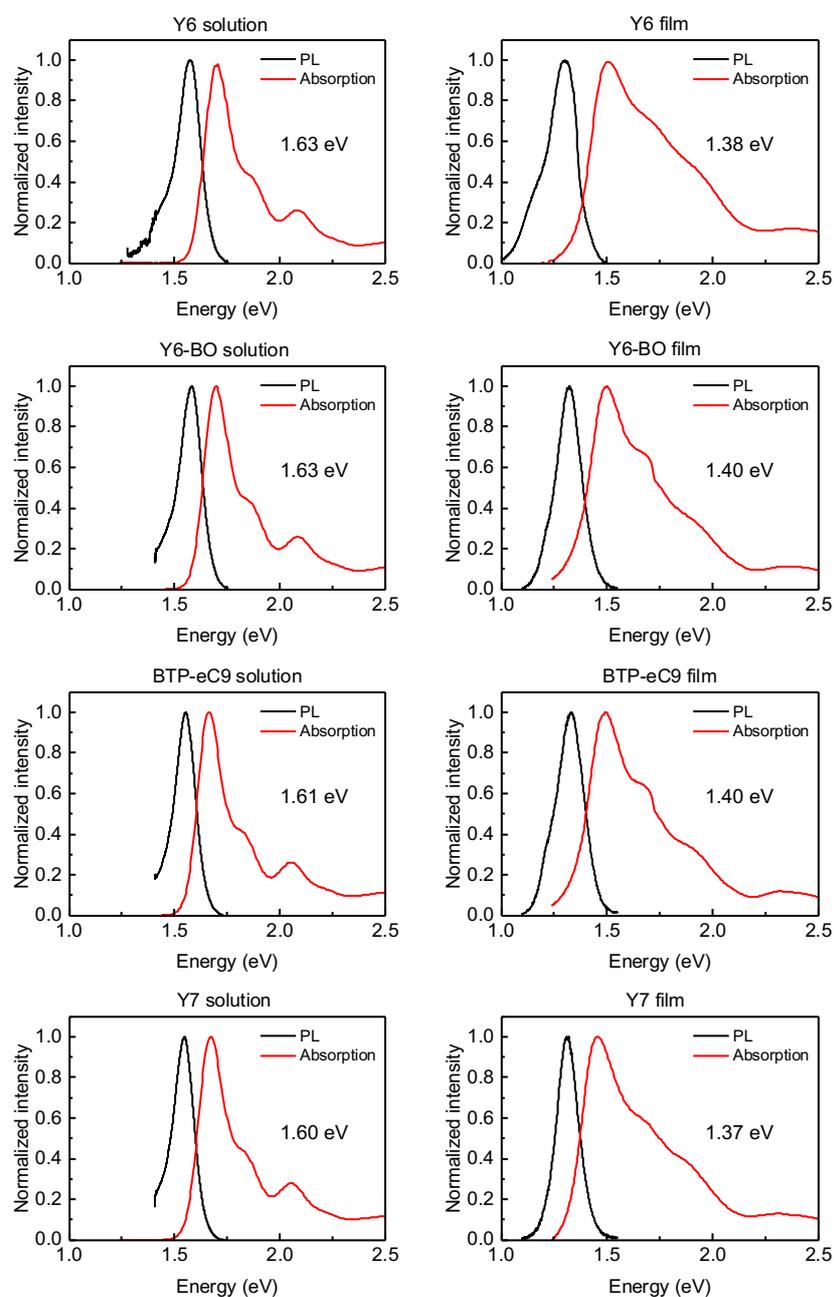

**Fig. S1.** Steady-state absorption and emission spectra of Y6-NFA systems (Y6, Y7, Y6-BO, BTP-eC9) in dilute solution (dissolved in chlorobenzene, $10^{-3}$ mg/ml) and in spin-coated thin-films. As described in main text, the intersection energies (shown in the plots) provide a measure of the material's optical gap (i.e. the first excited state energy, $E_{S1}$).

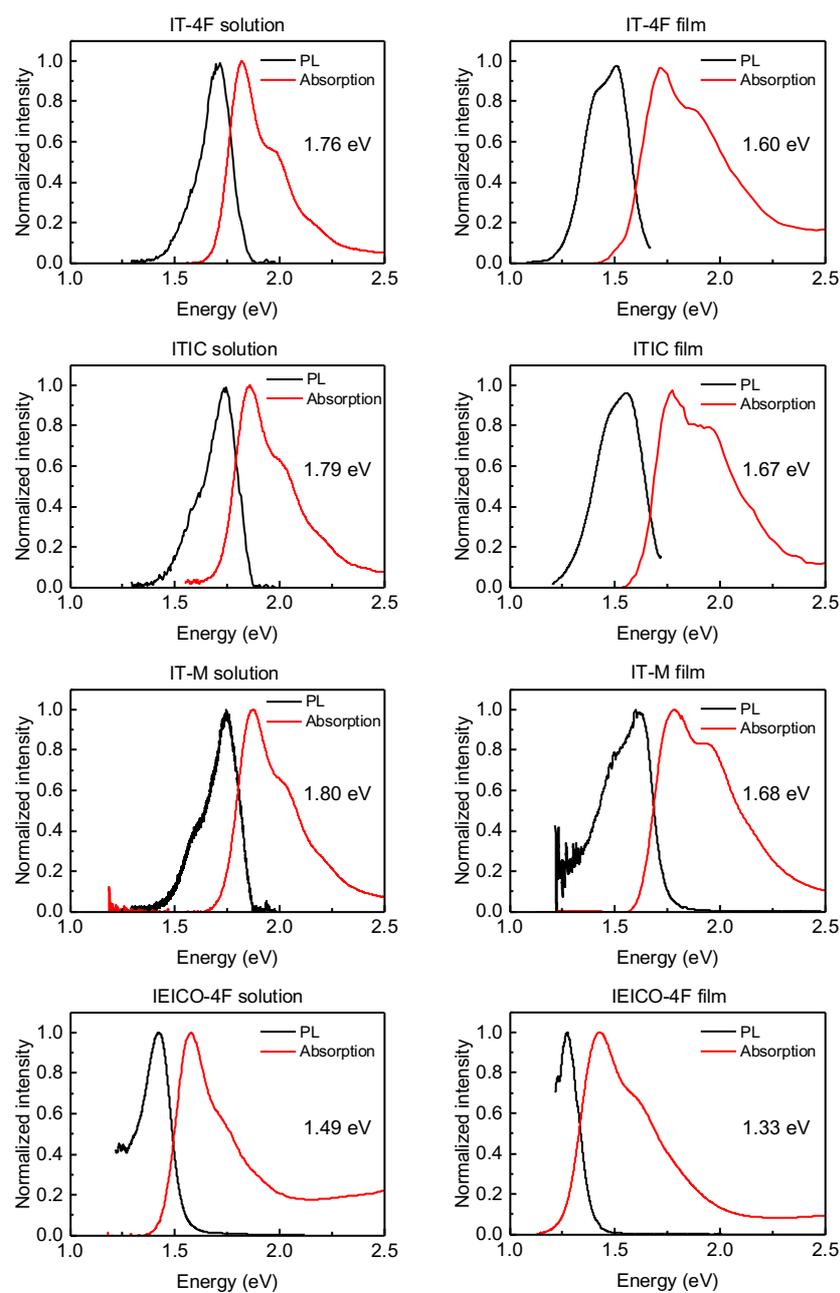

**Fig. S2.** Steady-state absorption and emission spectra of I-NFA systems (ITIC, IT-4F, IT-M, IEICO-4F) in dilute solution (dissolved in chlorobenzene, 10⁻³ mg/ml) and in spin-coated thin-films.

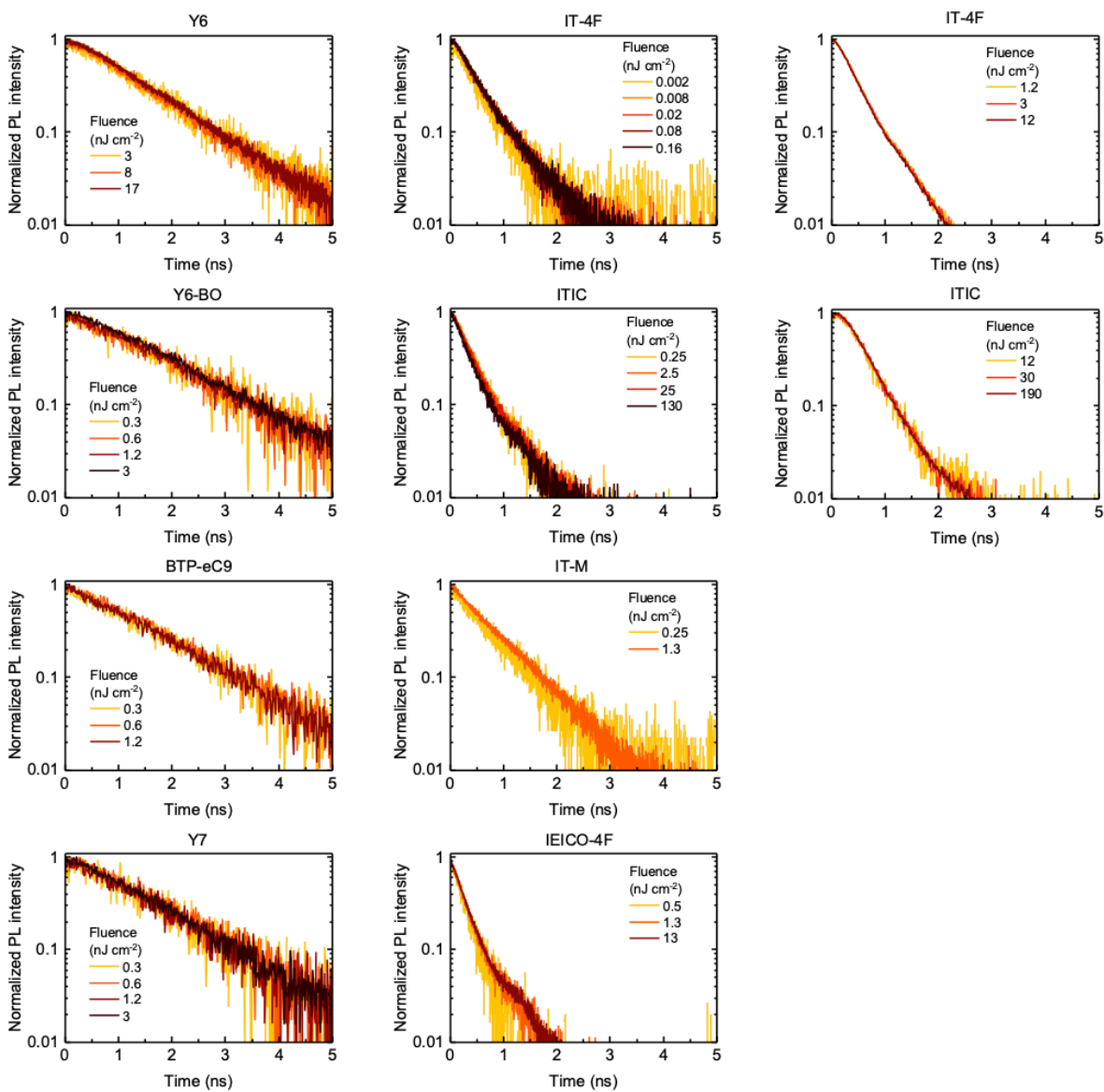

**Fig. S3.** Time-resolved photoluminescence (PL) of Y6-NFA and I-NFA film samples under various excitation laser intensities, showing negligible annihilation effects on the excited state lifetime under these experimental conditions. Results for dilute solution samples also show the same trend, as expected due to the negligible intermolecular interactions.

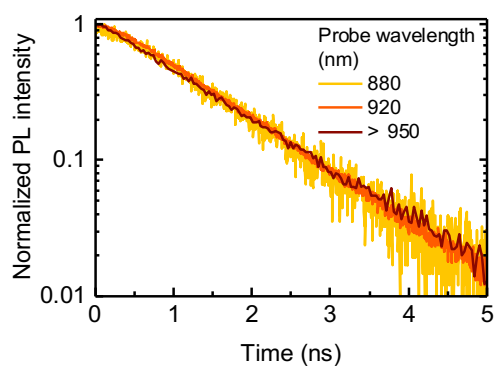

**Fig. S4.** Time-resolved photoluminescence (PL) of Y6 film probed at 880 nm (1.41 eV) and 920 nm (1.35 eV), measured using a silicon single photon detector. Data measured using an InGaAs single photon counter with a 950 nm long-pass filter is also shown for comparison. The results show very similar emission lifetime across the PL spectrum, thus indicating that the emissions arise from the same excited state.

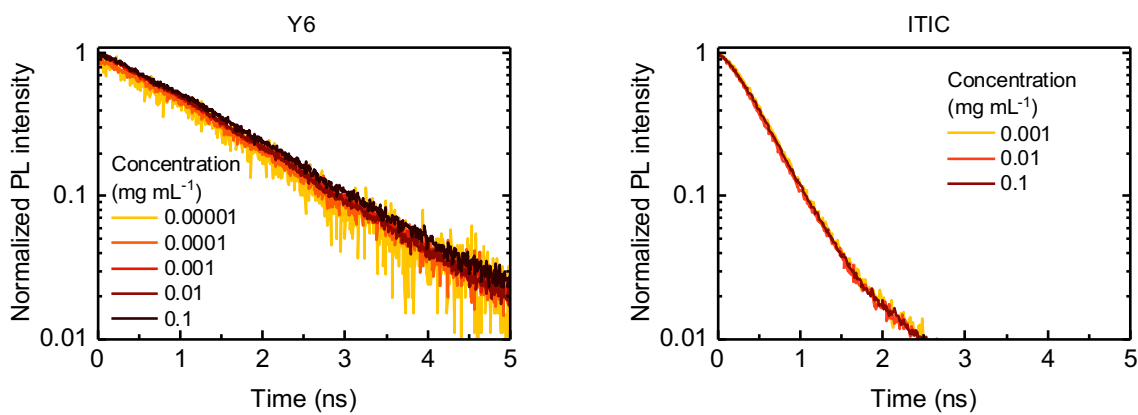

**Fig. S5.** Time-resolved photoluminescence (PL) of Y6 and ITIC solutions in chlorobenzene diluted to various concentrations, showing negligible change in emission lifetime across this range.

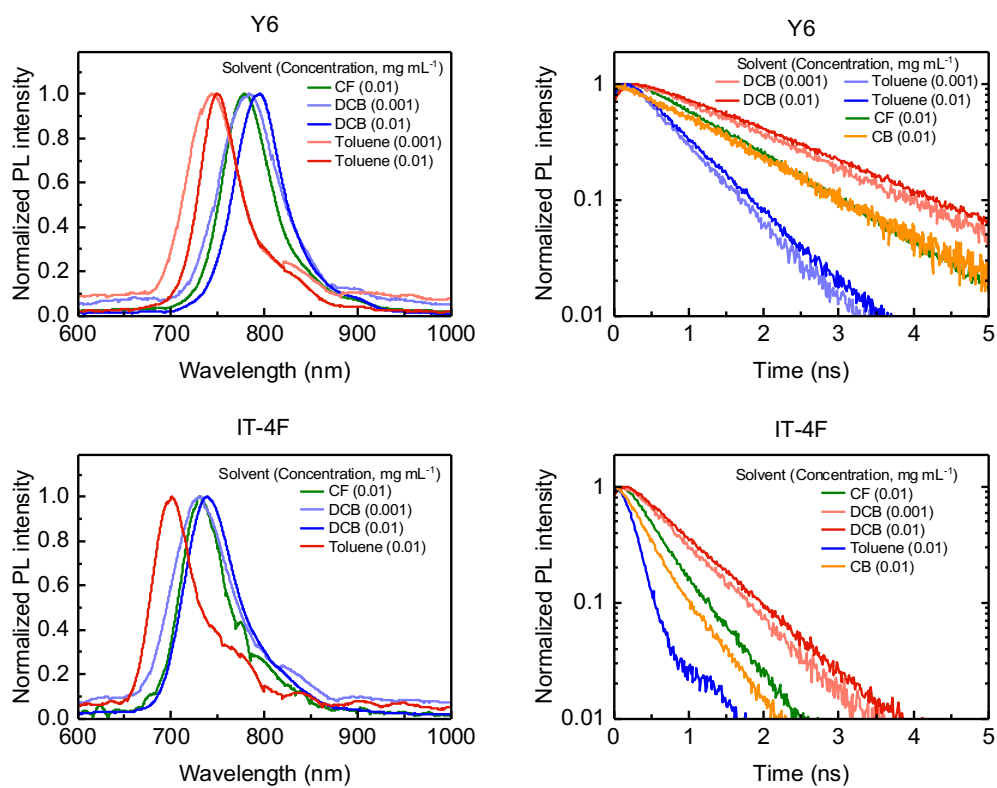

| Solvent | Concentration (mg/ml) | Dipole moment (Debye) | IT-4F solution PL lifetime | Y6 solution PL lifetime |
|---|---|---|---|---|
| Toluene | 0.01 | 0.31 | 0.27 ns | 0.71 ns |
| Chloroform | 0.01 | 1.15 | 0.48 ns | 1.25 ns |
| Chlorobenzene | 0.01 | 1.54 | 0.46 ns | 1.35 ns |
| Dichlorobenzene | 0.01 | 2.14 | 0.7 ns | 1.66 ns |

**Fig. S6.** Photoluminescence spectra and decay kinetics of Y6 and IT-4F dissolved in various solvents, namely toluene, chloroform (CF), chlorobenzene (CB) and dichlorobenzene (DCB). Both emission spectra and lifetimes showed solvent dependence, as expected due to the difference in dipole moments of the solvent (dipole moment values taken from: https://people.chem.umass.edu/xray/solvent.html). To ensure a systematic comparison between Y6-NFA and I-NFA in this study, all solution data presented in the main text were taken in chlorobenzene.

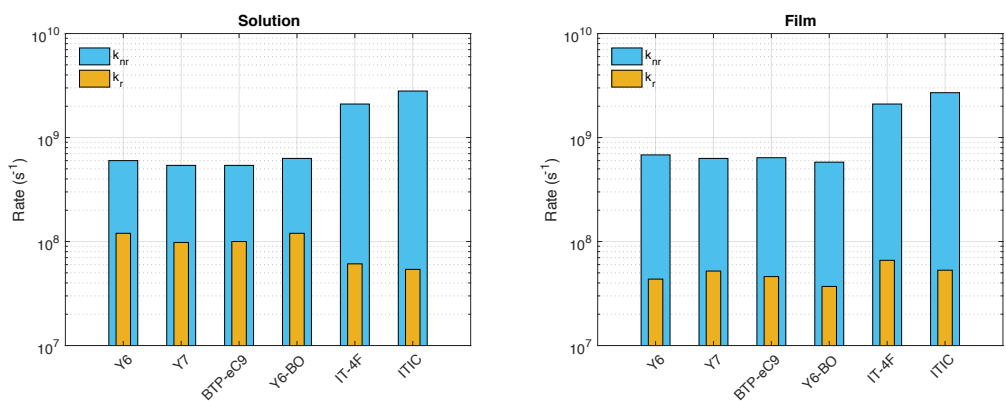

**Fig. S7.** Comparison of radiative ($k_r$) and non-radiative recombination ($k_{nr}$) rates for the various NFA materials in dilute solution (chlorobenzene) and in spin-coated thin-films. The same data is plotted in Fig. 2e in the main text. The rates were calculated based on the PLQY and PL lifetime measured by experiments, as described in the main text.

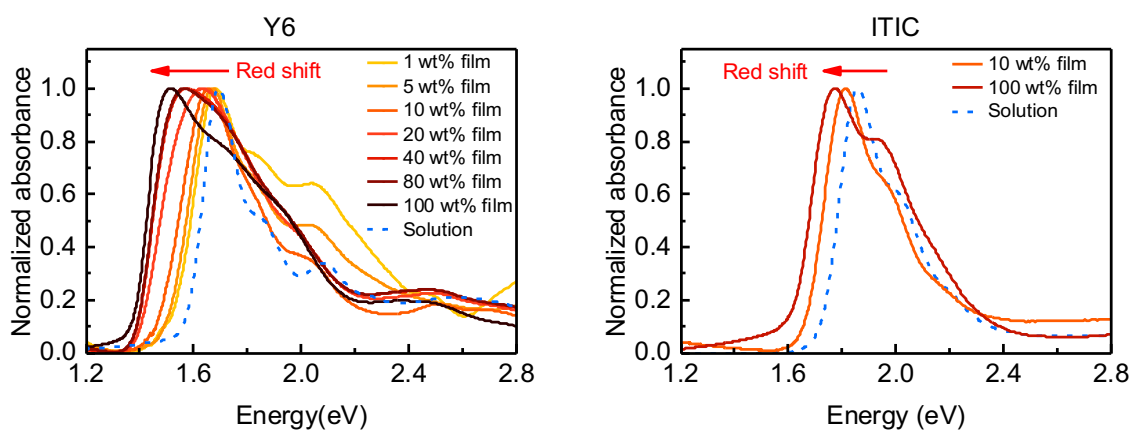

**Fig. S8.** Absorption spectra of dispersed Y6 and ITIC molecules in PVK spin-casted films. With low NFA to PVK weight ratio, the absorption spectrum matches with that measured in dilute solutions, thus indicating that the molecules are sufficiently dispersed for the electroabsorption measurements. The first and second derivative spectra shown in Figure 3 in the main text (blue and green lines, respectively) were determined from these absorption spectra.

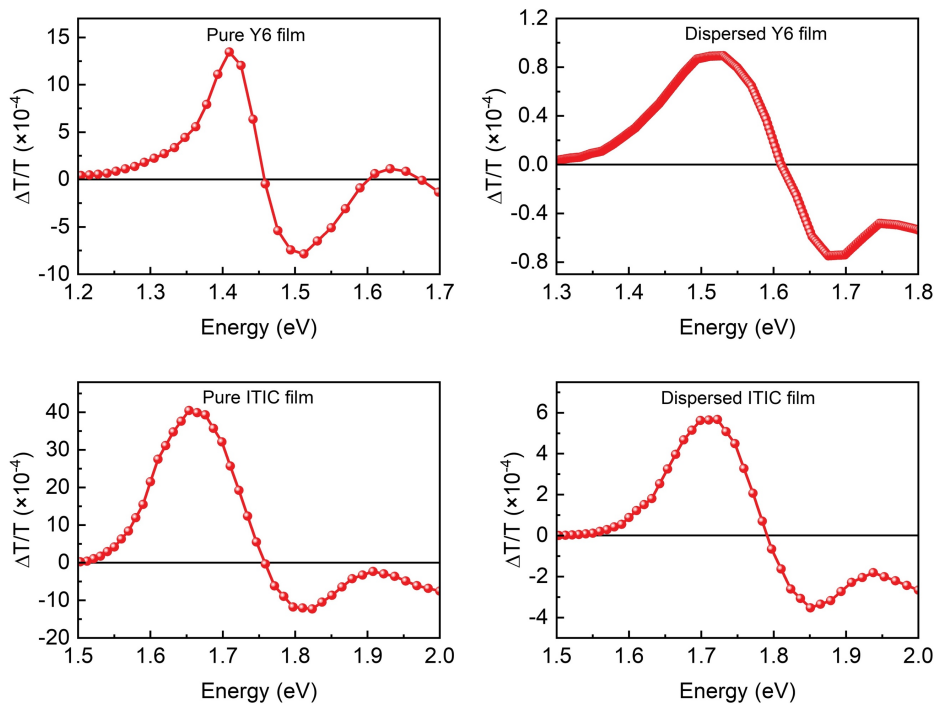

**Fig. S9.** Unnormalized electroabsorption data.

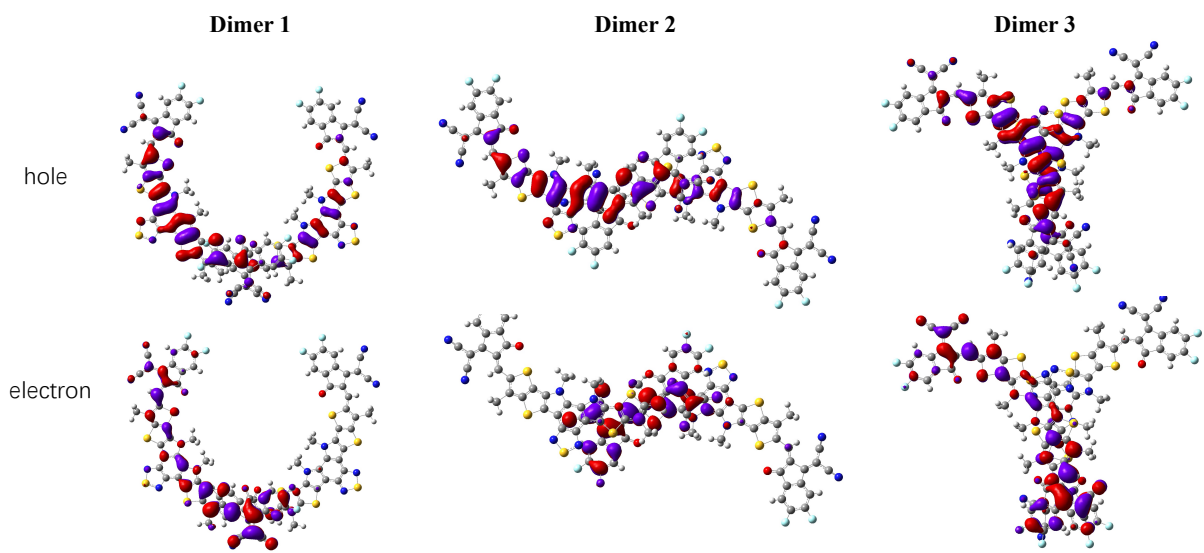

**Fig. S10.** Calculated Natural Transition Orbitals (NTO) describing the excitation character of the $S_1$ states in the three Y6 dimers (viewed from top). The corresponding side view NTOs, along with the calculated $S_1$ energies, hole/electron overlap and fluorescent oscillator strength are shown in Figure 4 of the main text.

| | | $E_{S1}$ (eV) | PLQY (± 0.5%) | PL lifetime in ns (± 0.5 ns) | $k_r$ (s$^{-1}$) | $k_{nr}$ (s$^{-1}$) |
|---|---|---|---|---|---|---|
| Y6 | Solution | 1.63 | 17.0% | 1.39 | $1.2 \times 10^8$ | $6.0 \times 10^8$ |
| | Film | 1.38 | 6.0% | 1.38 | $3.6 \times 10^7$ | $6.9 \times 10^8$ |
| Y7 | Solution | 1.60 | 15.3% | 1.56 | $9.8 \times 10^7$ | $5.4 \times 10^8$ |
| | Film | 1.37 | 7.6% | 1.46 | $5.2 \times 10^7$ | $6.3 \times 10^8$ |
| BTP-eC9 | Solution | 1.61 | 15.9% | 1.55 | $1.0 \times 10^8$ | $5.4 \times 10^8$ |
| | Film | 1.40 | 6.7% | 1.45 | $4.6 \times 10^7$ | $6.4 \times 10^8$ |
| Y6-BO | Solution | 1.63 | 15.9% | 1.33 | $1.2 \times 10^8$ | $6.3 \times 10^8$ |
| | Film | 1.40 | 6.0% | 1.63 | $3.7 \times 10^7$ | $5.8 \times 10^8$ |
| IT-4F | Solution | 1.76 | 2.8% | 0.46 | $6.1 \times 10^7$ | $2.1 \times 10^9$ |
| | Film | 1.60 | 3.1% | 0.47 | $6.6 \times 10^7$ | $2.1 \times 10^9$ |
| ITIC | Solution | 1.79 | 1.9% | 0.35 | $5.4 \times 10^7$ | $2.8 \times 10^9$ |
| | Film | 1.67 | 1.9% | 0.36 | $5.3 \times 10^7$ | $2.7 \times 10^9$ |
| IT-M | Solution | 1.80 | - | - | - | - |
| | Film | 1.68 | - | 0.69 | - | - |
| IEICO-4F | Solution | 1.49 | - | 0.85 | - | - |
| | Film | 1.33 | 0.6% | 0.39 | $1.54 \times 10^7$ | $2.5 \times 10^9$ |

**Table S1.** Summary of the excited state parameters determined by experiments following the procedures described in the main text.